\documentclass[preprint2]{aastex}

\usepackage{soul}
\usepackage{color}

\slugcomment{}

\shorttitle{Effects of thermohaline convection in A stars}
\shortauthors{Th\'eado et al.}

\begin{document}

\title{Influence of thermohaline convection on diffusion-induced iron accumulation in A stars}

\author{S. Th\'eado and S. Vauclair}
\affil{Laboratoire d'Astrophysique de Toulouse-Tarbes, Universit\'e de Toulouse, CNRS, 14 avenue Edouard Belin, 31400 Toulouse, France}
\email{stheado@ast.obs-mip.fr}
\author{G. Alecian}
\affil{LUTH, Observatoire de Paris, CNRS, Universit{\'e} Paris Diderot,
             5 Place Jules Janssen, 92190 Meudon, France}

\and

\author{F. LeBlanc}
\affil{D\'epartement de Physique et d'Astronomie, Universit\'e de Moncton, Moncton, NB, E1A 3E9, Canada}

\begin{abstract}
Atomic diffusion may lead to heavy element accumulation inside stars in certain specific layers. Iron accumulation in the Z-bump opacity region has been invoked by several authors to quantitatively account for abundance anomalies observed in some stars, or to account for stellar oscillations through the induced $\kappa$-mechanism. These authors however never took into account the fact that such an accumulation creates an inverse $\mu$-gradient, unstable for thermohaline convection. Here, we present results for A-F stars, where abundance variations are computed with and without this process. We show that iron accumulation is still present when thermohaline convection is taken into account, but much reduced compared to when this physical process is neglected. The consequences of thermohaline convection for A-type
stars as well as for other types of stars are presented.

\end{abstract}




\section{Introduction}

The importance of atomic diffusion \citep{Michaud70} inside stars is now well established: not only can it modify the atmospheric abundances, as observed in the so-called ``chemically peculiar stars'', but it can also have strong implications for the stellar internal structure. The variations with depth of the radiative accelerations on specific elements can lead to their accumulation or depletion in various layers inside the stars. In this framework, the accumulation of iron (together with less important metals like nickel) may have strong consequences, which are different according to the stellar type. 

In main sequence A stars, it may lead to the creation of an extra convective zone due to an iron-induced increase of opacity \citep{Richer00,Richard01}. If overshooting leads to complete mixing between this extra convective zone and the one due to helium, helium settling may be slowed down in the outer stellar layers. Such an effect could explain the existence of evolved oscillating Am stars \citep{Turcotte00}. The occurrence of an iron convective zone may also help explaining the excitation of g-modes in $\gamma$ Doradus stars, through the ``convective blocking'' process, as suggested by \citet{Guzik00}. Furthermore, in some cases iron accumulation can trigger stellar pulsations through the iron-induced $\kappa$-mechanism. This process has been invoked to explain main sequence B-type pulsators like SPB and $\beta$ Cephei stars \citep{Pamyatnykh04,Bourge06}. \citet{Charpinet96} predicted that the extreme horizontal branch stars referred to as sdB stars could also be destabilized due to iron-induced $\kappa$-mechanism. The oscillations of these stars were independently observed \citep[e.g.][]{Odonoghue97}. The qualitative success of the iron accumulation theory is impressive. However when quantitatively comparing models and observations, some difficulties remain. We will further discuss these difficulties in section \ref{discussion}.

In previous stellar modelling done in this framework, the question of the stability of a heavy iron layer lying above layers with smaller mean molecular weights was not addressed. This represents a serious problem, because of the existence of local inverse $\mu$-gradients, which are highly unstable against thermohaline (double diffusive) convection \citep[see][]{Vauclair04}. The induced mixing has to be taken into account in the computations, otherwise the results which are obtained do not correctly represent the real situation found inside the stars. 

The importance of thermohaline convection has been invoked in a few astrophysical cases. The first case which was pointed out concerned the $^3$He burning regions in giant stars \citep{Ulrich72,Charbonnel07}. Situations with $^4$He enhancement in outer stellar layers were also discussed in several studies. This enhancement could be due to mass transfer \citep{Stothers69} or to helium diffusion in a stellar wind \citep{Vauclair75}. The observations of main sequence helium-rich stars, with effective temperatures around 20 000K, show that helium is enhanced in average by a factor of two. This demonstrates that thermohaline mixing is important, otherwise the helium overabundance would be much larger. However, thermohaline convection mixing cannot be extremely efficient since this would lead to no helium overabundance. In this case, the reason for this lack of efficiency may be due to the presence of a magnetic field. The same kind of processes may occur inside roAp stars, as discussed by \citet{Balmforth01}.

Thermohaline convection has also been studied for the case of accretion of metal-rich matter onto main sequence stars. This situation may occur for exoplanet host stars, if they accrete hydrogen poor matter in the process of planetary formation. Because of thermohaline convection, and contrarily to preliminary conclusions found in early papers \citep{Santos01}, the accreted matter does not remain inside the convective zone, but diffuses downwards \citep{Vauclair04}. Another case concerns carbon enhanced main-sequence stars (CEMPs), which are assumed to have suffered some accretion of material coming from an AGB companion. This heavy matter falls down inside the star due to thermohaline convection \citep{Stancliffe07}. However, owing to the stabilizing $\mu$-gradient induced by helium diffusion, some of it can remain in the outer layers and may account for the observed abundances \citep{thompson08,Stancliffe08,Stancliffe09}. 

In this paper, we first give a general discussion of thermohaline convection in stars (section \ref{thermoconv}) and show how this physical process can be introduced in the stellar evolution computations (section \ref{computations}). Then results are presented for heavy elements accumulation inside A-type stars. We mainly concentrate on the evolution of a 1.7 $M_{\odot}$ star, but also give results for 1.5 $M_{\odot}$ and 1.9 $M_{\odot}$ stars (section \ref{results}). For these computations, the TGEC code \citep[cf.][]{Richard04,Hui08} was modified to include the computations of radiative accelerations using the SVP (standing for Single Valued Parameters) approximation proposed by \citet{Alecian02}. This modified version of the TGEC code was first used to compute stellar evolution models (without introducing thermohaline convection \citep{Theado09}) which were compared to the results obtained by the Montreal group. These computations were repeated while including thermohaline convection below the iron accumulation layer. Due to the stabilizing effect of helium settling, which fortunately occurs in the same stellar region, iron can still accumulate, but the abundance profiles are strongly modified. We also discuss the potential important consequences of this process for other types of stars in section \ref{discussion}. A short conclusion will then follow.

\section{Thermohaline convection}
\label{thermoconv}
Thermohaline convection is a well-known process in oceanography: salted water layers lying above fresh water ones are rapidly mixed downwards even in the presence of stabilizing temperature gradients, due to the different diffusivities of heat and salt. When a warm salted blob falls down in cool fresh water, the heat diffuses out more quickly than the salt. The blob goes on falling due to its weight until it mixes with the surroundings. This leads to the formation of so-called ``salt fingers''. 

A similar kind of convection occurs in stellar radiative zones in the presence of positive $\mu$-gradients, that is $\mu$-values increasing upwards. An``iron layer'' lying above hydrogen rich medium will lead to ``iron fingers'', because when a blob falls down heat diffuses between the blob and its surroundings more quickly than iron ions. With reference to oceanography, this process is called ``thermohaline'' convection. It is also sometimes referred to as double diffusive convection, because it is directly related to the presence of two “scalars” which diffuse on very different time scales. 

Another kind of double diffusive convection occurs in the reversed case, in the presence of destabilizing temperature gradients and stabilizing $\mu$-gradients like, for example, at the edge of convective cores. This process, quite different from thermohaline convection, is generally referred to as “semi convection”. It will not be considered here.

In the case of positive $\mu$-gradients, the medium
can become dynamically unstable if (Ledoux criterion):
\begin{equation}
\nabla_{crit} = \frac{\phi}{\delta}\nabla_{\mu} + \nabla_{ad} - \nabla < 0  
\end{equation}
where $\phi=(\partial$ ln $\rho/\partial$ ln $\mu)$ and $\delta=(\partial$ ln $\rho/\partial$ ln $T)$.
When $\nabla_{crit}$ vanishes, marginal stability is achieved. For positive or null values of $\nabla_{crit}$, the medium is stable against dynamical convection. Then thermohaline convection occurs on a time scale which is typically of a few thousand years, long compared to the dynamical time scales but short compared to stellar lifetimes. 

Fingers may form if the following condition is verified:
\begin{equation}
1 \leq |\frac{\delta (\nabla_{ad} - \nabla)}{\phi (\nabla_{\mu})}| 
\leq \tau^{-1} 
\end{equation}
with $\tau = D_{\mu} / D_T$  = $ \tau_T / \tau_{\mu}$ where $D_T$ 
and $D_{\mu}$ are the thermal and molecular diffusion
coefficients while 
${\tau_T}$ and $\tau_{\mu}$ are the corresponding time scales.

In stars the value of the $\tau$ ratio is typically on the order of $10^{-10}$ if one assumes that $D_{\mu}$ is the ``atomic'' diffusion coefficient. However, it can become larger when the shear flow instabilities which induce mixing between the edges of the fingers and the surroundings are taken into account \citep[see][]{Vauclair04}.

The effects of thermohaline convection as a mixing process in stars are far from trivial. Many detailed studies related to the occurence of this physical process in water have been published. For example, \citet{Gargett03} gave precise comparisons between numerical simulations and laboratory experiments. However the stellar case is different since mixing occurs in a compressible fluid. 

Two different parametrisation recipes, given respectively by \citet{Ulrich72} and \citet{Kippenhahn80} can differ by up to two orders of magnitude: this illustrates that treating thermohaline convection as a simple diffusion process may lead to inaccurate results. The basic problem here concerns the vertical shear flow instability which occurs between the fingers and the inter-fingers medium. This instability leads to local turbulence which increases the mixing at the edge of the fingers. Consequently, a process of self destruction appears for the blobs, so that fingers eventually reach a regime where they cannot form anymore: this effect is taken into account in the Kippenhahn et al. procedure.

In the present paper, we take into account the fact that the thermohaline mixing time scales are short compared to the main sequence evolution time scales, and that mixing stops when the positive $\mu$-gradients vanish. In our computations (see discussion below), we model rapid mixing so that the stellar medium always adjusts to keep either flat $\mu$-profiles or $\mu$-values increasing downwards. As helium settles downwards, it creates a stabilizing $\mu$-gradient which in turn may allow some heavy elements accumulation.

\section{Computations}
\label{computations}
We computed stellar models to describe the evolution of A-type stars while using the Toulouse-Geneva Evolution Code (TGEC). The code is described in detail in \citet{Richard04} and \citet{Hui08}, except for the computations of opacities and atomic diffusion which have undergone major improvements \citep{Theado09}.

\subsection{Opacities and radiative accelerations}
In the present computations, important improvements were introduced in the TGEC code concerning the opacities and radiative accelerations. The opacities were computed using the OPCD v3.3 codes and data\footnotemark \footnotetext{The OPCD\_3.3 packages is available on the following website: http://cdsweb.u-strasbg.fr/topbase/op.html} \citep{Seaton05}. It allows to compute self consistent Rosseland opacities taking into account the detailed composition of the chemical mixture. The opacities were recalculated at each time step.

To more realistically treat atomic diffusion in the TGEC code, the radiative accelerations on C, N, O, Ca and Fe were included following the improved version \citep{Leblanc04} of the semi-analytical prescription proposed by \citet{Alecian02}.
This method allows for very fast computations of radiative accelerations with a reasonable accuracy. The SVP approximation may be implemented in existing codes in a simple way. There are much less data to process than for detailed radiative acceleration calculation, because complete and detailed monochromatic opacites for each ion are not needed. A new grid of SVP-parameters, well fitted to the stellar mass range considered in this work, was computed following the procedure described in \citet{Leblanc04}.

The diffusion of H, He and the five metals mentioned above were included in the stellar models. The abundance of the other elements were assumed to be constant throughout the star. The radiative accelerations of these five heavy elements as well as the detailed abundances for all seven elements were computed at each diffusion time step, chosen to be about 50 times smaller than the evolution time step.

\subsection{Convection and Mixing}
For our modelling, the convective regions were instantaneously homogenised. The HI and HeII convective regions were supposed close enough to be connected by overshooting and mixed together. On the other hand, the iron convective region which may appear in some models in much deeper regions, was supposed disconnected from the surface convective zone.

The models were evolved from pre-main sequence up to hydrogen core exhaustion. They were assumed homogeneous on the pre-main sequence and atomic diffusion was introduced at the beginning of the main sequence.

We first computed models without thermohaline convection. To avoid the appearance of steep and unrealistic abundance gradients at the transition between radiative and convective regions, we introduced mild mixing at the bottom of each convective zone. This mixing was modeled as a diffusion process through a coefficient D$\rm _{turb}$ introduced in the chemical transport equation:
\begin{equation}
\rho \frac{\partial \bar{c_i}}{\partial t}=\frac{1}{r^2}\frac{\partial}{\partial r} \left( r^2 \rho D_{turb} \frac{\partial \bar{c_i}}{\partial t} \right)
\label{transport}
\end{equation}
$\rm \bar{c_i}$ represents the mean concentration of a species i, $\rho$ is the density, r is the radius, D$\rm _{turb}$ is given by:
\begin{equation}
D_{turb}=D_{czb} \exp \left( \ln 2 \frac{r-r_{czb}}{\Delta} \right)
\end{equation}
D$\rm _{czb}$ and r$\rm _{czb}$ are respectively the value of D$\rm _{turb}$ and the value of the radius at the base of the convective zone, $\Delta$ is the half width of the mixing region. D$\rm _{czb}$ and $\Delta$ are free parameters, choosen to produce a mild mixing on a small extent. The value of D$\rm _{czb}$ is taken equal to $10^5$cm$^2$.s$^{-1}$ and $\Delta$ is fixed to 0.5\% of the stellar radius below the surface convective region and to 0.05\% of the stellar radius below the iron convective zone.

\subsection{Modelling thermohaline convection}

As discussed in the previous sections, thermohaline convection arises in regions where the mean molecular weight increases towards the surface. It leads to mixing on short time scales, compared to those of evolution, and it disappears when the mean molecular weight gradient vanishes. To simulate this phenomenon, we found that the most efficient and simple way was to model the mixing as usual, provided that the chosen diffusion coefficient is large enough to flatten the $\mu$-profile any time a destabilizing $\mu$-gradient is created (at each time step). 

The diffusion coefficient proposed by \citet{Kippenhahn80} is convenient in this framework:
\begin{equation}
D_{th}=\frac{H_p}{\nabla_{ad}-\nabla} \frac{16 a c T^3}{c_p \kappa \rho^2} |\frac{d \ln \mu}{dr}|
\end{equation}
where $\rm H_{P}$ is the pressure scale height, a the radiative pressure, c the speed of light, T the temperature, $\nabla_{ad}$ the adiabatic gradient, $\nabla$ the real temperature gradient, $\rm c_P$ the specific heat at constant pressure, $\kappa$ the opacity and $\mu$ the mean molecular weight of the material. 

We used this prescription in our computations and checked that the $\mu$-gradient remained always null or negative at each time step in all our models.

\section{Results}
\label{results}
We computed models for 3 different masses: 1.5, 1.7 and 1.9M$_{\odot}$. Each set of models includes a first series of computations without thermohaline convection and a second series including it. 
The initial metal mixture used in these computations is the solar mixture presented by \citet{Grevesse93} with the following values as H and He initial mass fractions: $\rm X_0=0.7112$ and $\rm Y_0=0.2714$. The initial iron mass fraction was consequently $\rm X_0(Fe)=0.1149 \times 10^{-2}$.

\begin{figure}
\center
\includegraphics[width=8cm]{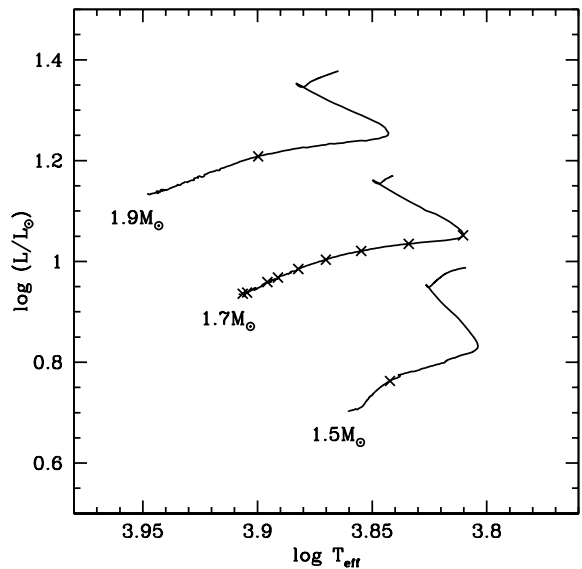}
\caption{Evolutionary tracks for three different masses, including atomic diffusion and thermohaline convection. The crosses locate the position of the models shown in Figures \ref{fe} to \ref{dlnmu}. Those situated on the 1.7 M$_\odot$ track locate models at 62 Myrs, 86 Myrs, 299 Myrs, 403 Myrs, 588 Myrs, 788 Myrs, 988 Myrs, 1188 Myrs and 1388 Myrs. On the 1.5 and 1.9 M$_\odot$
track, only two models are presented, respectively at 838 Myrs and 632
Myrs.}

\label{diaghr}
\end{figure}

Figure \ref{diaghr} presents the evolutionary tracks for the 3 sets of models. Inclusion of thermohaline convection does not modify these tracks in a sensible way: the two curves coincide. The crosses show the positions of the models which are discussed below.

\subsection{Results obtained for 1.7M$_{\odot}$ models with and without thermohaline mixing}

\begin{figure*}
\center
\includegraphics[width=\textwidth,bb=55 215 520 645]{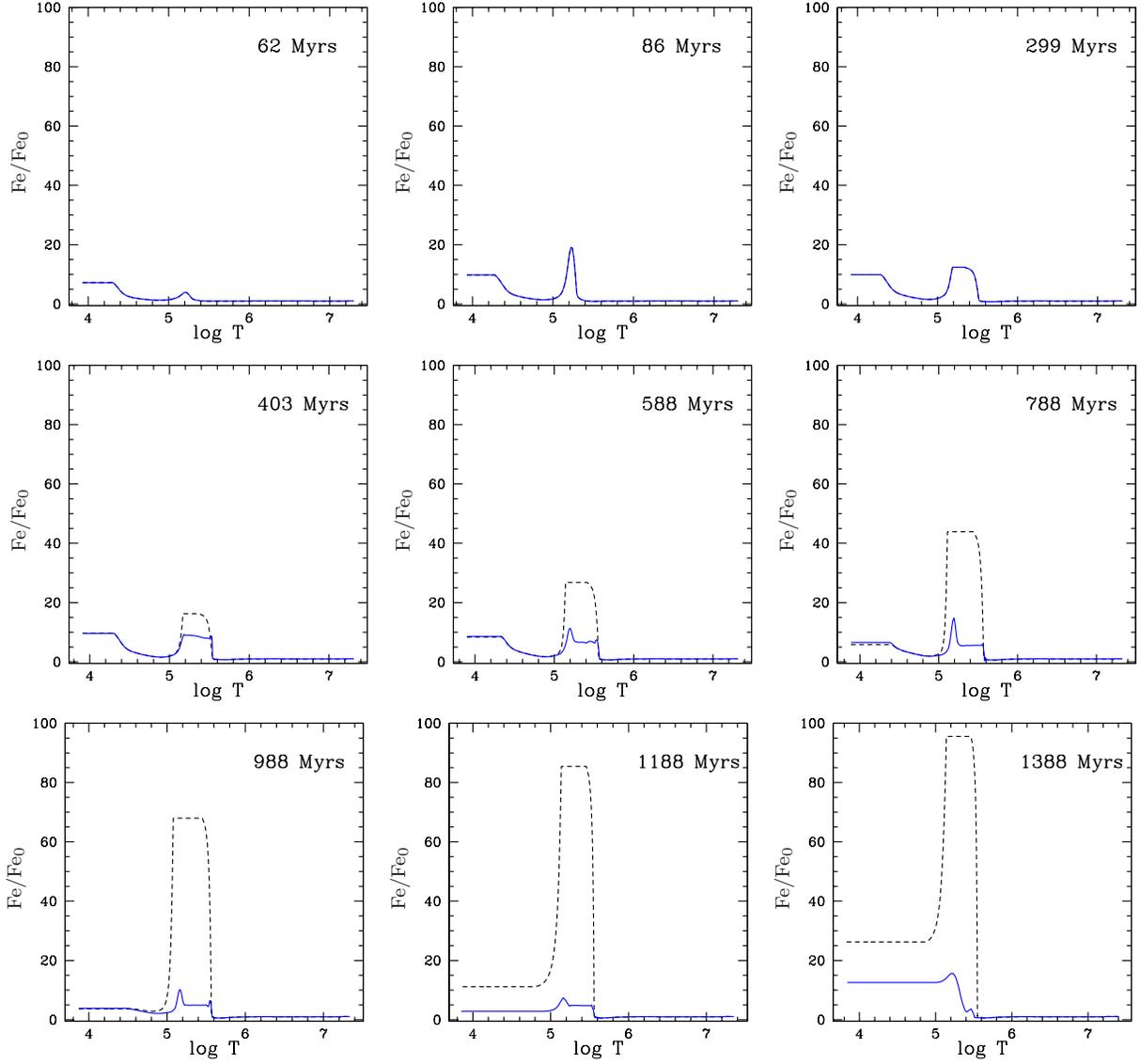}
\caption{Iron accumulation in two 1.7 M$_{\odot}$ models with atomic diffusion at various evolutionary steps. $\rm Fe/Fe_0$ represents the ratio between the iron mass fraction and its initial value. The dashed lines are for models without thermohaline convection and the solid lines for models including it.}
\label{fe}
\end{figure*}

\begin{figure*}
\center
\includegraphics[width=\textwidth,bb=55 215 520 645]{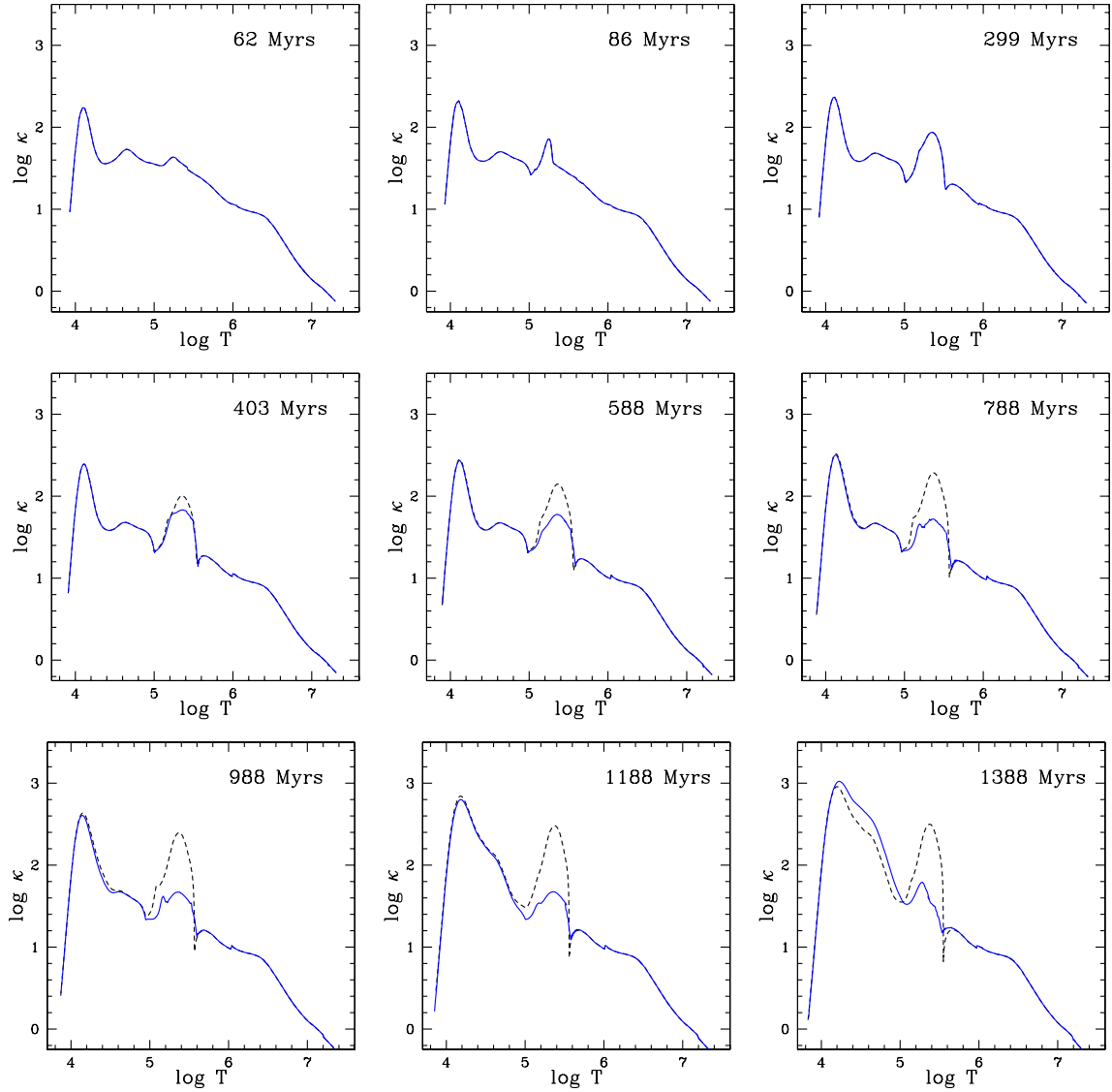}
\caption{Opacity profiles in the two 1.7 M$_{\odot}$ models presented in Figure \ref{fe}. Dashed lines: without thermohaline convection; solid lines: with thermohaline convection.}
\label{logk}
\end{figure*}

\begin{figure*}
\center
\includegraphics[width=\textwidth,bb=55 215 520 645]{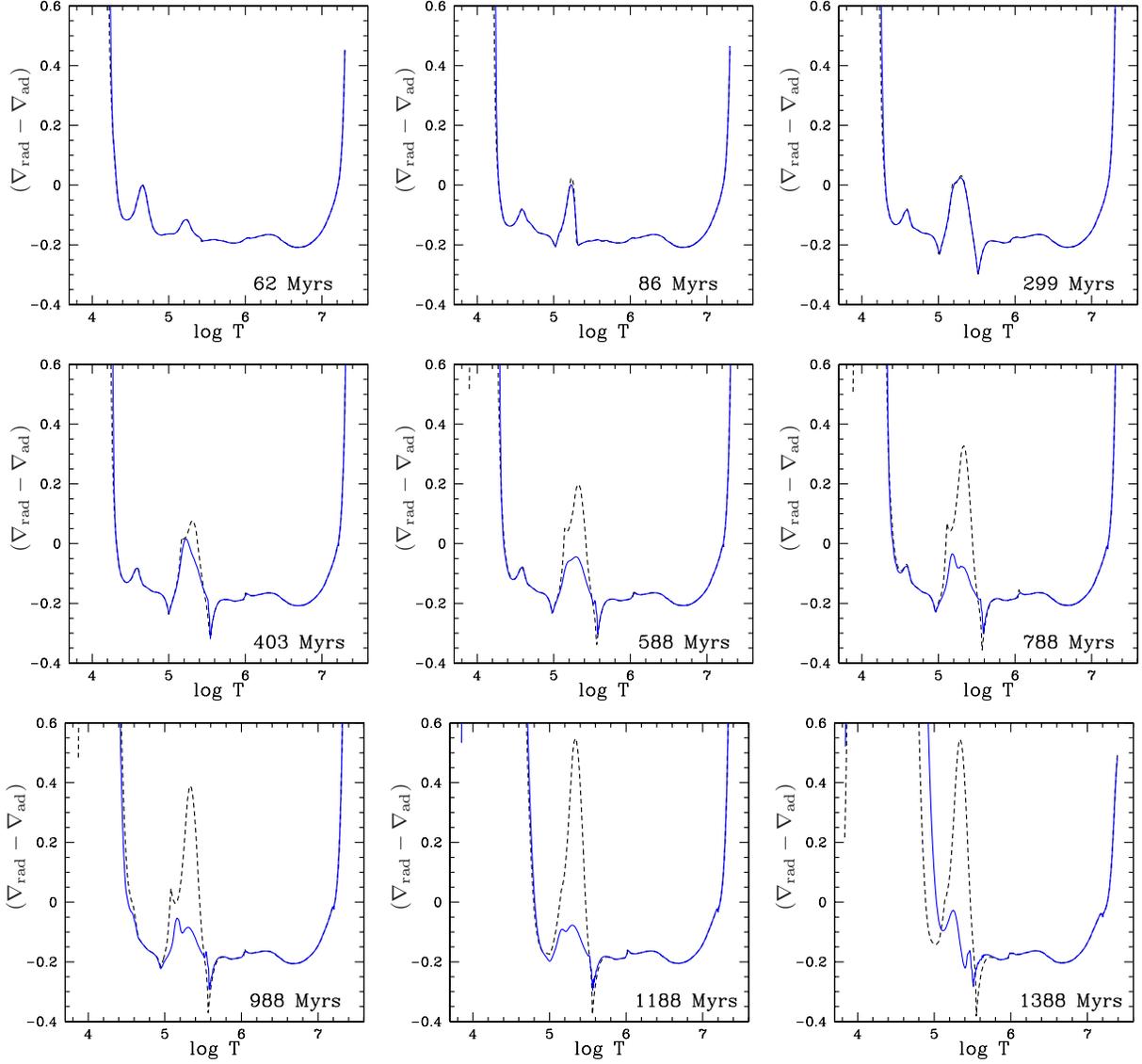}
\caption{Difference of the radiative and adiabatic gradients in the 1.7 M$_{\odot}$ models presented in Figure \ref{fe}. Convective regions appear when this difference is positive. Dashed lines: without thermohaline convection; solid lines: with thermohaline convection.}
\label{conv}
\end{figure*}

\begin{figure*}
\center
\includegraphics[width=\textwidth,bb=55 215 520 645]{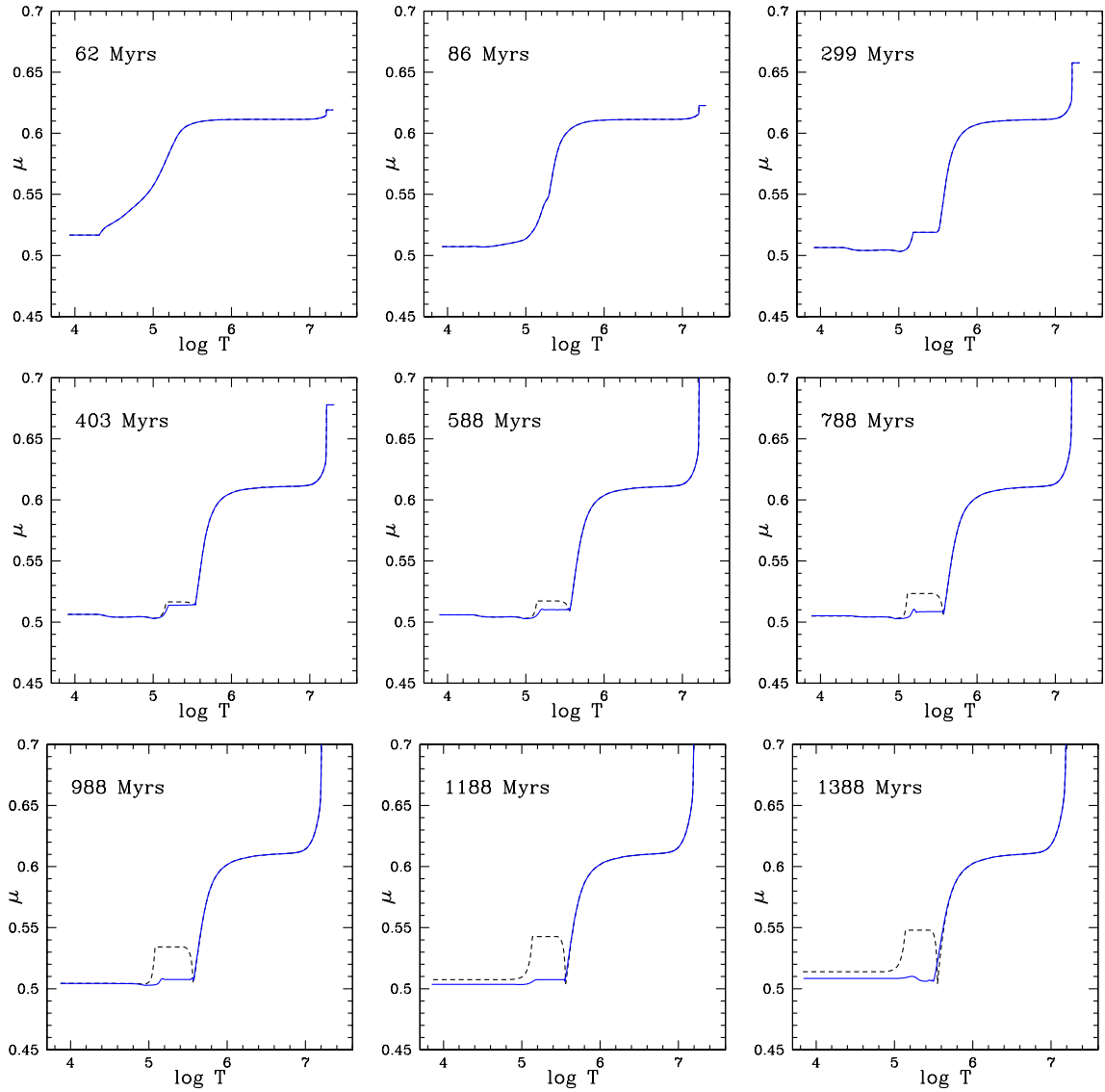}
\caption{Molecular weight profiles in the 1.7 M$_{\odot}$ models presented in Figure \ref{fe}. Dashed lines: without thermohaline convection; solid lines: with thermohaline convection.}
\label{mu}
\end{figure*}

\begin{figure*}
\center
\includegraphics[width=\textwidth,bb=55 215 520 645]{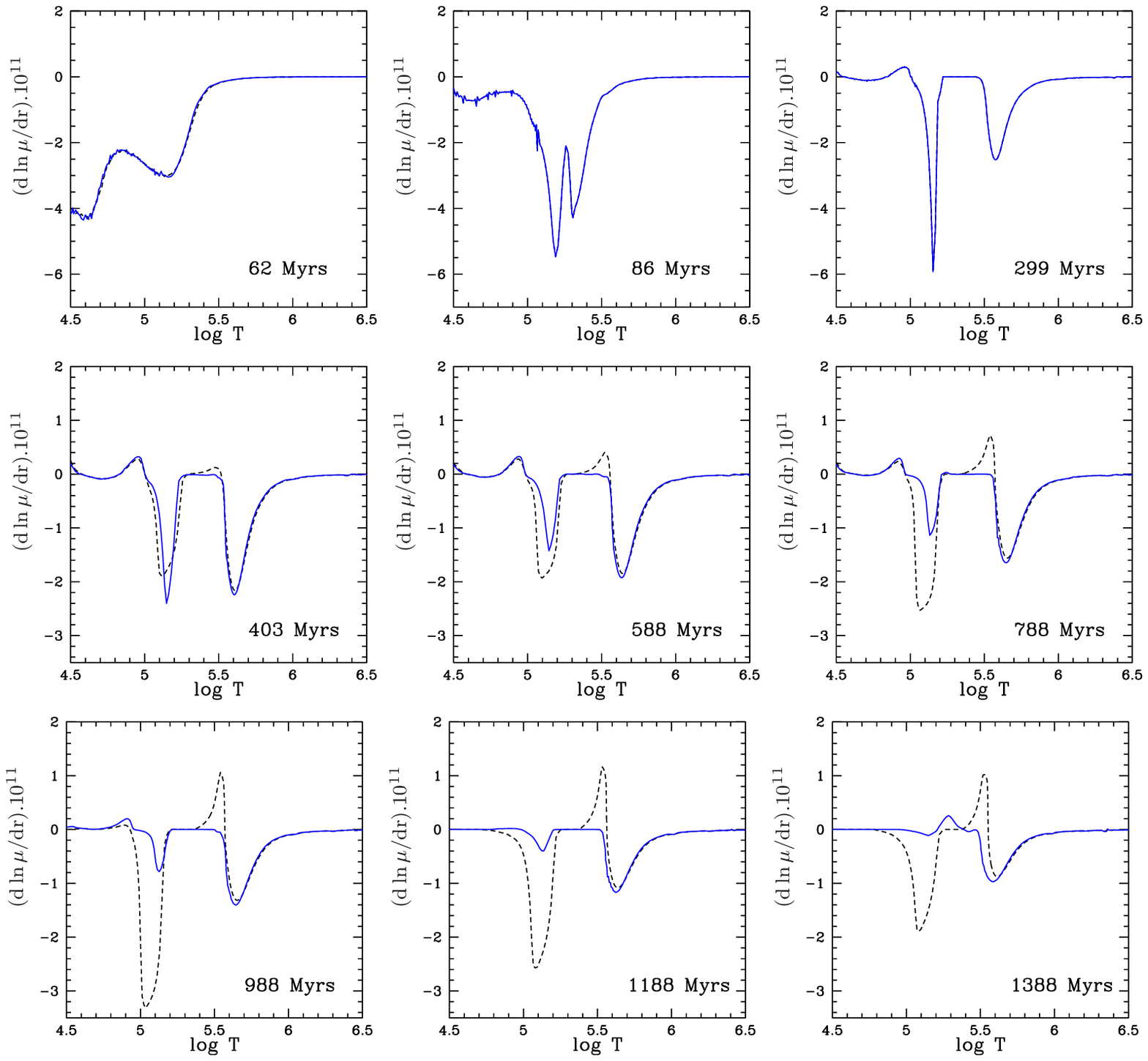}
\caption{Molecular weight logarithmic gradients in the 1.7 M$_{\odot}$ models presented in Figure \ref{fe}. Dashed lines: without thermohaline convection; solid lines: with thermohaline convection.}
\label{dlnmu}
\end{figure*}

For the discussion here, we specifically focus on the case of 1.7M$_{\odot}$ models. Figures \ref{fe} to \ref{dlnmu} compare the results obtained with (solid lines) and without (dashed lines) thermohaline mixing, at various epochs during the H burning phase (crosses in Fig. \ref{diaghr}). Figure \ref{fe} displays the ratio of the iron mass fraction at various times to that of its initial value. Figure \ref{logk} shows the opacity profiles at these same evolution times. Figure \ref{conv} presents the differences between the radiative and adiabatic gradients, which allows to locate the convective regions. Figure \ref{mu} displays the mean molecular weight profiles and Figure \ref{dlnmu} displays the mean molecular weight gradient, for the interval $ \rm 4.5 \le log T \le 6.5$.

The models reach the main sequence with chemically homogeneous envelopes. Then, due to atomic diffusion, helium sinks and creates a stable (negative) $\mu$-gradient below the convective envelope, whereas the radiative acceleration on iron induces an increase of its abundance in the upper stellar layers, and more specifically in the iron-peak elements opacity bump centered at $\rm \log T \simeq 5.2$. For the 1.7M$_{\odot}$ case, the models with and without thermohaline mixing begin to diverge at about 400 Myrs.

At the age of 62 Myrs, in both cases the iron accumulation reaches a factor of 7 below the surface, which does not alter the opacity, and a factor 4 in the opacity bump region, which leads locally to a small opacity increase. Consequently, the radiative gradient ($\displaystyle \nabla_{rad}=3/(16 \pi a c G). (P \kappa L_r)/(T^4 M_r)$) increases, but it remains smaller than the adiabatic gradient so that the corresponding region is still convectively stable. At this time (62 Myrs) the model only has a single surface convective zone, due to the ionization of hydrogen. As can be seen in Figure \ref{conv}, from the close-to-zero value of $(\nabla_{rad}-\nabla_{ad})$ around $\rm \log T \simeq 4.6$, the HeII convective zone has just disappeared, due to He settling.   
At this point, the iron accumulation is not large enough to significantly alter the $\mu$-profile. Here, the effect of the He-induced $\mu$-gradient is indeed much more important than the iron contribution. During stellar evolution, as the atomic diffusion proceeds, the region where the helium-induced stable $\mu$-gradient is the steepest slowly deepens inside the star. As soon as it occurs below the iron opacity bump, the effects of the iron accumulation on the $\mu$-profiles become important. 

At 86 Myrs, iron accumulation exceeds a factor of 19 and affects the $\mu$-profile. It is responsible for the change in the slope of the $\mu$-profile at $\log T \simeq 5.3$. The iron enrichment leads to a significant opacity increase and therefore to an increase of the radiative gradient which now exceeds the adiabatic one. In both cases, with or without thermohaline mixing, a new very narrow convective region appears around  $\rm \log T \simeq 5.2$.
 
At 299 Myrs, this iron convective region becomes significant and leads to chemical homogenization between $\rm \log T \simeq 5.2$ and 5.4 as seen in the $\mu$- and $(\rm Fe/Fe_0)$-profiles shown. The homogenization of the iron convective zone leads to an expansion of the iron accumulation region and reduces the iron maximum value from a factor of 19 to a factor of 12.5. From the beginning of the main-sequence until 299 Myrs, the $\mu$-gradient remains null or negative in most models, except below the surface convective zone where it sometimes takes a slightly positive value, due to iron enrichment in the outer regions. 

After 400Myrs, the $\mu$-gradient induced by He-settling is then located below the iron accumulation region and the effects of iron accumulation become visible. From now on, models including thermohaline convection below the iron accumulation layers significantly differ from those without thermohaline convection.

During the subsequent stellar evolution, radiative levitation leads to iron accumulation in the iron opacity bump region. In models without thermohaline convection, this iron accumulation drastically increases with time, reaching a factor of 20 at 400 Myrs and up to a factor of 95 at 1388 Myrs (Fig. \ref{fe}, dashed lines). The consequences on the opacity profiles can be seen in Figure \ref{logk}, and those on the radiative gradients in Figure \ref{conv}. It also leads to a new convective zone which persists during most of the main sequence lifetime. 

Iron accumulation in these specific layers leads to a spectacular increase of the $\mu$-values (Fig. \ref{mu}, dashed lines). After 400 Myrs, ``$\mu$-bumps'' develop with a rapid increase slightly above $\rm \log T \simeq 5.2$ and a steep decrease below $\rm \log T \simeq 5.4$, which is highly unstable. The corresponding $\mu$-gradients are shown in Figure \ref{dlnmu}.    

In the models including thermohaline convection induced by these unstable $\mu$-gradients, iron accumulation in the opacity bump region is drastically reduced and never exceeds a factor of 15. It is never completely suppressed however, due to the stabilizing effect of helium settling. The consequences on the opacity profiles, radiative gradients, $\mu$-values and their gradients can be seen in Figure \ref{logk} to \ref{dlnmu}, solid lines. The evolution of the internal stellar structure in the two cases studied here is sensibly different. 

Contrarily to the models without thermohaline mixing, those which include it do not have persistent iron convective zones, although marginal convection appears and disappears several times during stellar evolution. Actually, in models including thermohaline mixing, the iron abundance in the opacity bump always stays close to the critical value for the onset of convection. When iron accumulation exceeds this critical value, convective mixing occurs and reduces the iron accumulation, making this region stable again. Note that when this occasional convective zone appears, it has a smaller extent than the iron bump itself, so that only part of the iron enriched zone is homogenized. This is the reason why a small iron peak remains around $\rm \log T \simeq 5.2$ in the 58 Myrs, 788 Myrs and 988 Myrs models.

\subsection{Comparisons between the 1.5, 1.7 and 1.9M$_{\odot}$ cases}
\begin{figure*}
\center
\includegraphics[width=12cm,bb=110 254 465 605]{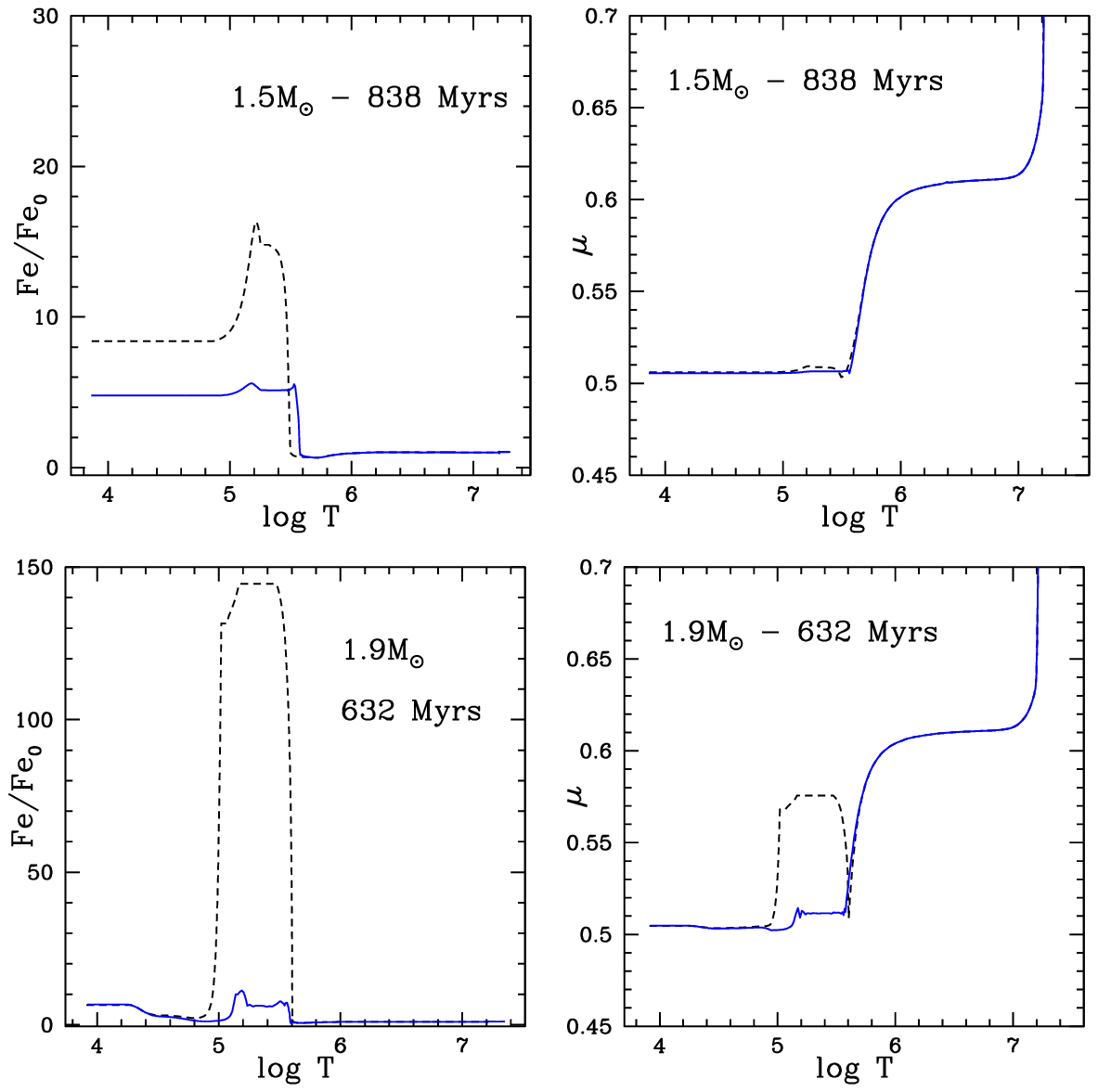}
\caption{Iron profiles (left panels) and $\mu$-profiles (right panels) in two 1.5 and 1.9M$_{\odot}$ main sequence models (with a central hydrogen mass fraction $\rm X_c\simeq 0.5$) including atomic diffusion. Dashed lines: without thermohaline convection; solid lines: with thermohaline convection.}
\label{compar3}
\end{figure*}

The results obtained for 1.5 and 1.9M$_{\odot}$ models are qualitatively similar to those of the 1.7M$_{\odot}$ models. To illustrate the efficiency of thermohaline convection in these models, we present in Figure \ref{compar3} the molecular weight and the iron profiles obtained with and without thermohaline mixing for one model for each mass, located at the middle of the main-sequence (cf. Fig. \ref{diaghr}). The radiative acceleration on iron increases with the stellar mass, so that, when no thermohaline convection is taken into account, the iron accumulation is larger in more massive models. The consequent ``$\mu$-bump'' is also more prominent, as can be seen in Figure \ref{compar3} (dashed lines). In the results obtained including thermohaline mixing (solid lines), we can see that iron accumulation is again drastically reduced. In the 1.9M$_{\odot}$ model, it does not exceed a factor of 10 whereas it goes up to 150 when thermohaline mixing is neglected.

\begin{figure*}
\center
\includegraphics[width=\textwidth,bb=55 280 520 575]{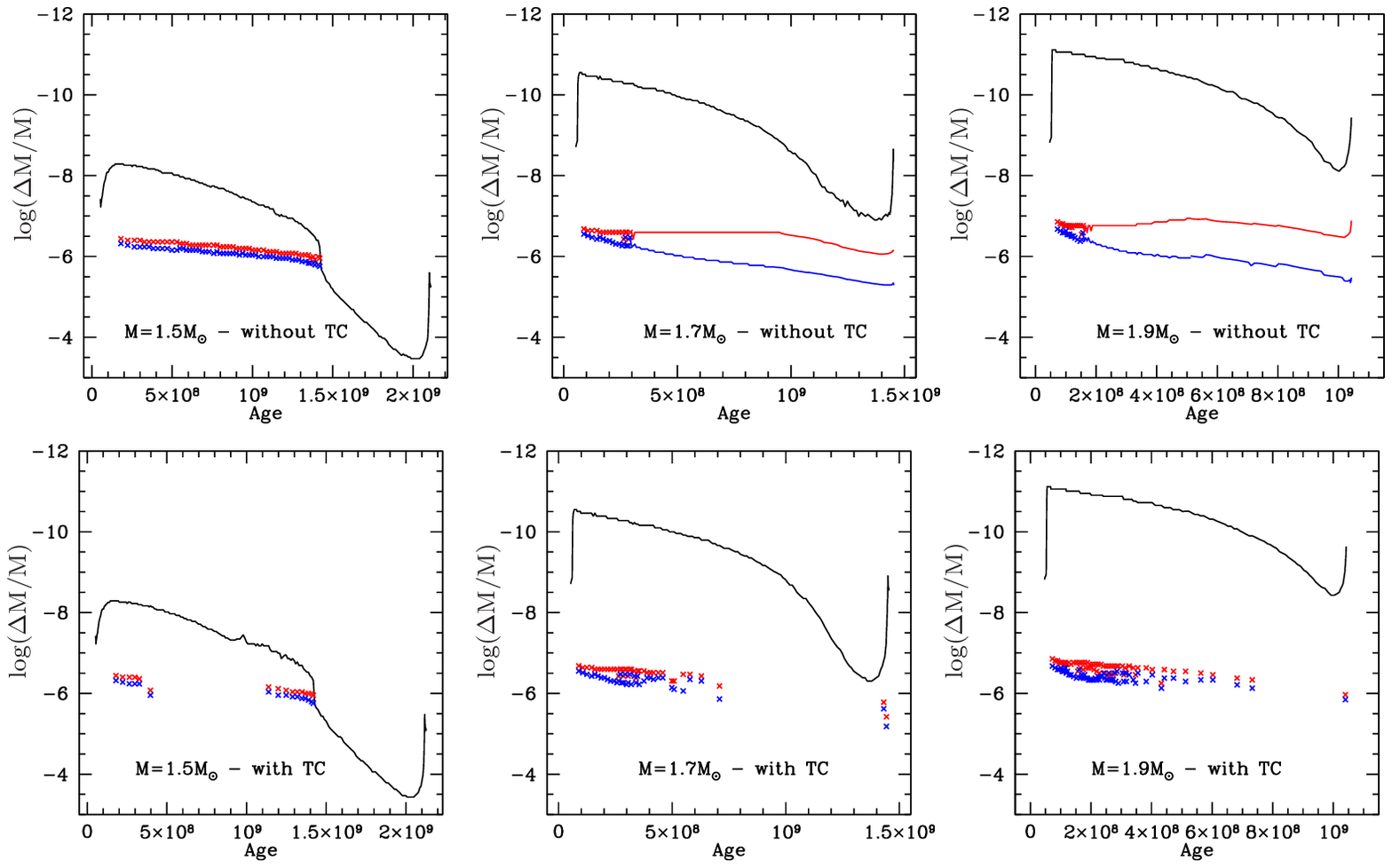}
\caption{Convective zones in models with (lower panels) and without (upper panels) thermohaline mixing. The vertical axis, $\displaystyle \rm \log (\Delta M/M)$, represents the outer mass fraction. For each plot, the upper solid lines locates the bottom of the surface (H-He) convective zone. During stellar evolution, a new convective zone induced by iron accumulation appears deeper in the stellar interior (typically around $-7< \log (\Delta M/M) <-5$). In some models this iron convection zone settles durably while other models undergo mixing episodes followed by convectively stable phases. During the mixing episodes, the boundaries of the iron convective zone are represented by crosses. During the convectively unstable phases, they are represented by solid lines.}
\label{zconv}
\end{figure*}

Figure \ref{zconv} presents the position of the boundaries of the convective zones inside the models with and without thermohaline mixing for the three masses studied here. The crosses locate the boundaries of the iron convective zone during phases with mixing episodes. When the convective zone settles durably, its boundaries are represented with solid lines. 

For the three stellar masses considered, an iron convective zone resulting from a diffusion-induced iron accumulation rapidly appears after the beginning of the main sequence. In models without thermohaline mixing (Fig. \ref{zconv}, upper panels), a first phase occurs during which convective episodes are followed by convectively stable periods. When convection appears in the opacity bump region, the iron peak decreases due to convective mixing. Consequently the opacity also decreases and the radiative gradient becomes smaller than the adiabatic gradient. The region becomes radiatively stable again. In the 1.5M$_{\odot}$ models, this succession of convective and radiative episodes goes on until the end of the main sequence. In the 1.7 and 1.9M$_{\odot}$ models, after a few hundred megayears, iron accumulation has become large enough (more than a factor of 15), so that convective mixing cannot reduce the iron abundance enough to stop convection. The iron convective zone then persists during the rest of the main sequence lifetime.  

In models including thermohaline convection, the situation is different. Here again, there is a first phase during which mixing episodes alternate with convectively stable periods. Afterwards the onset of thermohaline mixing strongly modifies the convective (un)stability of the iron opacity bump region. Iron accumulation is reduced and leads, most of the time, to a radiative gradient slightly below the adiabatic gradient. When, during stellar evolution, the accumulation of iron exceeds, somewhere in the opacity bump region, the critical value for the onset of convection, the induced mixing rapidly reduces the iron abundance and the radiative stability is restored. As a result the iron convective zone does not persist in these models. 

\section{Discussion}
\label{discussion}
Thermohaline convection is expected to take place inside stars any time inverted $\mu$-gradients are built. This happens in all types of stars where heavy element accumulation occurs in specific stellar layers, either due to radiative levitation or to nuclear reactions during stellar evolution. It also occurs in case of accretion of hydrogen poor material onto stellar surfaces, due to planet swallowing in early phases, or to accretion of matter from a companion. Here we presented how introducing thermohaline convection modifies the results obtained for the accumulation of iron due to diffusion inside A-type stars. Some other cases where this process has to be introduced were already discussed in previous papers (e.g. \citet{Charbonnel07,Stancliffe07}). Before giving our general conclusions in the next section, we first propose a discussion of three other cases where thermohaline convection should be introduced and for which we expect important consequences. 

\subsection{The SPB / $\beta$ Cephei stars}
The SPB (slowly pulsating B stars) and the $\beta$ Cephei stars are main sequence B type pulsators. The SPB stars oscillate with periods between 0.5 to 5 days, while the $\beta$ Cephei stars possess 2 to 8 hours-oscillations. The former are interpreted as high-order g-modes, the latter as low order p and/or g-modes.
In the early nineties, the new OPAL opacity determinations \citep{Rogers92} allowed to solve the long standing enigma of B type pulsators. Rosseland mean opacities computated from these tables were substancially increased in the Z-bump region compared to the previously computed opacities. This opacity enhancement (as large as a factor of 3) provided a boost for the $\kappa$-mechanism potentially activated in the metal opacity bump region and could then explain the excitation of pulsations in $\beta$ Cephei and SPB theoretical standard models \citep{Cox92,Kiriakidis92,Moskalik92,Dziembowski93b}. However, as observational capabilities progress, B type pulsators offer new challenges to the theorists. Some of them are now found in low metallicity environments \citep[e.g.][and references therein]{Kolaczkowski06}. A large number of pulsation modes are indeed detected in B stars, with frequency spectra sometimes revealing new discrepancies between theory and observations. As an example, the two $\beta$ Cephei stars, $\nu$ Eri and 12 Lac \citep[][and references therein]{Handler06,Jerzykiewicz05} present frequency spectra which cannot be reproduced with B type stars standard models \citep[e.g. for $\nu$ Eri,][]{Ausseloos04}. \citet{Pamyatnykh04} showed that an ``ad hoc'' factor 4 enhancement of the iron group elements, in the Z-bump region could help reproducing the observed pulsation spectrum of $\nu$ Eri. The combined effects of radiative levitation and gravitional settling provides a natural explanation for this local metal enrichment and could then explain the observation of more than 60 $\beta$ Cephei stars in low metallicity environments.
 
Following the results of \citet{Pamyatnykh04}, \citet{Bourge06} presented preliminary evolutionary models of $\beta$ Cephei stars evolved from the ZAMS up to 1 Myr: their evolutionary models did not include consistent diffusion computations but they included, at each timestep, a rough estimation of the diffusion-induced iron accumulation in the Z-bump region. After one million years, their inferred iron enrichment is increased by a factor of 12 in the iron opacity bump. The stability analysis of these models showed that such iron enrichments makes the theoretical frequency spectra denser, which could help reducing the discrepancies between asteroseismic observations and theoretical stability analysis. 

Following the same idea, \citet{Miglio07b} carried out a parametric study of the effects of local iron enhancement on the stability of SPB and $\beta$ Cephei stars. The parametric iron accumulation profile introduced in their models is described by a gaussian function centered at $\rm \log T \simeq 5.2$, calibrated to be consistent with the iron accumulation profile deduced by \citet{Richard01} in their A and F star models. As \citet{Bourge06}, \citet{Miglio07b} showed that such accumulation profiles widen the theoretical SPB and $\beta$ Cephei instability strips and increases the number of excited modes.

The computations by \citet{Bourge06} and \citet{Miglio07b} reduced the discrepancies between theoretical and observed pulsations. However they did not discuss the credibility of the iron profiles introduced in their models, nor the stability of the induced $\mu$-gradients.

On the other hand, \citet{Miglio07a} analysed the effects of uncertainties in the opacity computations on the excitation of pulsation modes in B type stars. They compared models computed with OPAL opacities \citep{Iglesias96} with models computed with the recently updated OP opacity tables \citep{Seaton05,Badnell05}. The OP updated data lead to an enhancement of 18\% of the opacity in the Z-bump. \citet{Miglio07a} showed that these differences considerably affect the theoretical instability strips of $\beta$ Cephei and SPB stars. Using the OP opacities drastically increases the number of excited modes and significantly extends the theoretical instability strips. These new opacity computations could help solve, at least partly, the discrepancies between theoretical predictions and observations in $\beta$ Cephei, SPB and hybrid stars. Up to now, no evolutionary models have been computed for B-type stars including consistent diffusion computations and the new OP opacities. However in the light of \citet{Miglio07a} results, we expect the iron accumulation derived by \citet{Pamyatnykh04} to recouncile observations and theory (by a factor of 4) to be overestimated. In this context the iron enrichment inferred by \citet{Bourge06} (by a factor of 12) for a 1 Myr-model should also be too large, which suggests the presence of a ``forgotten'' process (such as the thermohaline convection) which could reduce the diffusion-induced iron accumulation.

\subsection{The sdB stars}
Contrarily to SPB stars, subdwarf B stars (sdB stars) are evolved, compact objects with low-mass cores and outer H-rich envelopes. They are experiencing the core helium burning phase on the extended horizontal branch. They all present chemical peculiarities attributed to gravitational settling and radiative levitation in the presence of weak stellar winds. Subdwarf B stars host two groups of pulsators: the first one, composed of hot sdB stars, is characterised by rapid ocillations, with periods ranging from 80 to 600s, caused by low-order, low-degree p-modes. The second group, composed of the coolest sdB stars, oscillates with periods in the 2000-9000s range, due to high-order, low-degree g-modes. Both types of pulsators are driven by the $\kappa$-mechanism acting in the iron-peak element opacity bump.

The relation between the envelope metal content and the pulsation driving mechanism was first established by \citet{Charpinet96} for the short-period B subdwarf pulsators. They showed that a solar metal content in the envelope of their models is unable to excite modes but that models with envelopes containing an overabundance of metals can globally destabilize modes such as observed. They argued that this metal enrichment needs only to occur in the driving region itself and not necessarily in the whole envelope as assumed in their models. 

A natural explanation for the required local metal enhancement is the radiative levitation process, which is expected to act efficiently in the radiative envelopes of subdwarf B stars. In these stars, as in A and F stars, the competition between gravitational settling and radiative aceleration is expected to produce local accumulations of heavy elements. \citet{Charpinet97} presented more sophisticated and realistic models of sdB stars, the so-called ``second generation models'', including non uniform iron abundances. These envelope models rely on the assumption that a state of diffusive equilibrium is reached between radiative levitation and gravitational settling operating on iron (disregarding other potentially competing processes), leading to stable non uniform iron profiles. \citet{Charpinet97} demonstrated that such local iron accumulations in the Z-bump region may produce efficient excitation for pulsation modes. Similar models were successfully used for detailed asteroseismic studies of several sdB stars: they globally nicely reproduced the global properties of the pulsating sdB stars and were able to closely reproduce their observed periods \citep{Brassard01,Charpinet01,Charpinet05a,Charpinet05b,Charpinet06,Charpinet08}.

As emphasized by \citet{Charpinet09}, these ``second generation models'' however suffer from several shortcomings.
The theoretical instability strip predicted by these models is larger than the observed one and the observed period ranges in individual stars are also usually narrower than predicted. These discrepancies are likely due to an overestimate of the iron abundance in the Z-bump region.

In the ``second generation models'' of Charpinet et al., the opacities are computed using the OPAL opacity tables \citep{Iglesias96}. As discussed previously for SPB and $\beta$ Cephei stars, the use of the most recent OP opacities \citep{Seaton05,Badnell05} may also have a direct impact on the properties of the sdB pulsation driving mechanism. Similarly to other types of stars \citep{Jeffery07,Miglio07a}, using OP opacities instead of OPAL data, would increase the opacity in the Z-bump, resulting in a more efficient pulsation driving mechanism. 

Here again we find evidences that the iron accumulation needed to account for the observations must be lower than obtained by the atomic diffusion alone. Introducing thermohaline convection could certainly help reducing the discrepancies.

\subsection{The $\gamma$ Doradus stars}
The $\gamma$ Doradus stars are A-F main sequence pulsators. Their oscillations are interpreted as high-order g-modes with periods ranging from 0.35 to 3 days. They are driven by a flux blocking mechanism at the base of their convective envelope, as shown by \citet{Guzik00} with frozen convection models and by \citet{Dupret04} and \citet{Dupret05} with time-dependent convection models. It has been demonstrated that the ``convection blocking'' of radiation can drive high order g-modes only in stellar models for which the temperature at the bottom of the convective envelope, ranges between $2 \times 10^5$K (which corresponds nearly to the Z-bump region central temperature) and  $4.8 \times 10^5$K. The depth of the convective envelope plays a crucial role in the driving mechanism proposed by \citet{Guzik00} to explain the $\gamma$ Doradus pulsations. 

As a consequence, the seismic properties of $\gamma$ Doradus stars are quite sensitive to atomic diffusion including radiative levitation because of the way they influence the convective zones. As emphasized by \citet{Montalban07}, the diffusion effects on convection are of various kinds. The He settling which leads to H-enrichment in the external stellar layers induces an opacity increase which affects the depth of the surface convective zone. On the other hand, as discussed in previous paragraphs, radiative levitation produces reservoirs of iron-peak elements which may induce the appearance of an iron convective zone near $2 \times 10^5$K. Consequently the driving mechanism of $\gamma$ Doradus stars is expected to be sensitive to the thermohaline convection which occurs below the heavy elements accumulation zone. 

\section{Conclusion}

In the present paper, we tested the effect of thermohaline convection on the amplitude of iron accumulation which occurs inside A-F type stars due to atomic diffusion. We computed two series of models for three stellar masses: 1.5, 1.7 and 1.9M$_{\odot}$. In both series, atomic diffusion was consistently computed for the elements H, He, C, N, O, Ca and Fe. The computations included the radiative accelerations on C, N, O, Ca and Fe, by using the semi-analytical prescription proposed by \citet{Alecian02} and improved by \citet{Leblanc04}. 

No thermohaline convection was introduced in the first series of models. We verified that our results were similar to those obtained by \citet{Richer00} and \citet{Richard01}, with a large accumulation of iron in the iron peak elements opacity bump region. Thermohaline convection was then introduced for the second series of models. Considering the short time scales of this mixing process compared to the evolution time scales, and the fact that mixing occurs until the unstable $\mu$-gradients disappear, we introduced it in such a way that the $\mu$-gradients were kept close to zero or negative all along the evolutionary tracks. In practice, we used an extra diffusion coefficient, as proposed by \citet{Kippenhahn80}, which is large enough to reach this purpose.

We presented in detail the results obtained for the 1.7M$_{\odot}$ case, during the main sequence phase. We also showed some results for the two other stellar masses considered here. The effect of thermohaline convection is very efficient in strongly reducing the heavy element accumulation in the opacity bump region. Whereas the iron abundance can increase by up to factors larger than 100 when thermohaline convection is neglected, it never increases by more than a factor of 15 when it is taken into account. 

It is important to note that some iron accumulation always remains, even in the presence of thermohaline convection. The reason why this happens is directly related to helium settling, which leads to a stabilizing $\mu$-gradient. This stabilizing effect allows in turn some heavy material abundance increase up to the point where the $\mu$-profile becomes flat.

As a consequence, the cases where iron accumulation has been invoked to account for specific observations in stars remain. Indeed, in many cases the observations would be better explained with a smaller iron accumulation than that given by atomic diffusion alone. All these cases (Am stars, SPB stas, sdB stars, $\beta$ Cephei stars and $\gamma$ Doradus stars as discussed in Section 5) will have to be studied individually for a detailed comparison with the observations.

In the present computations, nickel was not included. Although less important than iron, it may significantly contribute to the opacity and will have to be included in future work. No mass loss nor extra turbulence were taken into account. Including such macroscopic motion may reduce the heavy element accumulation even without thermohaline convection. However, thermohaline convection is a physical process which must occur in any case every time the $\mu$-gradient is unstable. The final accumulation of elements thus strongly depends on this process, even in the presence of other macroscopic motion.

In this paper, we have demonstrated the importance of thermohaline convection which has to be taken into account in the computation of the abundance variations induced by atomic diffusion. Precise applications to specific stellar cases still have to be undertaken. They will lead to important consequences on the abundance anomalies observed in stars, as well as their oscillating properties through several processes like $\kappa$-mechanism, convective blocking, etc. The possible applications of thermohaline convection in stellar astrophysics are vast and will certainly be a topic of interest in years to come.

\acknowledgments


\begin{thebibliography}{}
\bibitem[Alecian \& LeBlanc(2002)]{Alecian02} Alecian, G., \& LeBlanc, F. 2002, \mnras, 332, 891
\bibitem[Ausseloos et al.(2004)]{Ausseloos04} Ausseloos, M., Scuflaire, R., Thoul, A., \& Aerts, C. 2004, \mnras, 355, 352
\bibitem[Badnell et al.(2005)]{Badnell05} Badnell, N. R., Bautista, M. A., Butler, K., Delahaye, F., Mendoza, C., Palmeri, P., Zeippen, C. J., \& Seaton, M. J. 2005, \mnras, 360, 458
\bibitem[Balmforth et al.(2001)]{Balmforth01} Balmforth, N. J., Cunha, M. S., Dolez, N., Gough, D. O., \& Vauclair, S. 2001, \mnras, 323, 362
\bibitem[Bourge \& Alecian(2006)]{Bourge06} Bourge, P.-O., \& Alecian, G. 2006, ASP Conference Series, vol. 349, 201 
\bibitem[Brassard et al.(2001)]{Brassard01} Brassard, P., Bill\`eres, M., Charpinet S., Liebert, J., \& Saffer, R. A. 2001, \apj, 563, L1013
\bibitem[Charbonnel \& Zahn(2007)]{Charbonnel07} Charbonnel, C., Zahn, J.P., 2007, \aap, 476, L29
\bibitem[Charpinet et al.(1996)]{Charpinet96} Charpinet S., Fontaine, G., Brassard, P., \& Dorman, B.  1996, \apj, 471, L103
\bibitem[Charpinet et al.(1997)]{Charpinet97} Charpinet S., Fontaine, G., Brassard, P., Chayer, P., Rogers, F. J., Iglesias, C. A., \& Dorman, B. 1997, \apj, 483, L123
\bibitem[Charpinet et al.(2001)]{Charpinet01} Charpinet, S., Fontaine, G., \& Brassard, P. 2001, \pasp, 113, 775
\bibitem[Charpinet et al.(2005a)]{Charpinet05a} Charpinet S., Fontaine, G., Brassard, P., Green, E. M., \& Chayer, P. 2005a, \aap, 437, 575
\bibitem[Charpinet et al.(2005b)]{Charpinet05b} Charpinet S., Fontaine, G., Brassard, P., Bill\`eres, M., Green, E. M., \& Chayer, P. 2005b, \aap, 443, 251
\bibitem[Charpinet et al.(2006)]{Charpinet06} Charpinet S., Silvotti, R., Bonanno, A., Fontaine, G., Brassard, P., Chayer, P., Green, E. M., Bergeron, P., et al. 2006, \aap, 459, 565
\bibitem[Charpinet et al.(2008)]{Charpinet08} Charpinet S., Van Grootel, V., Reese, D., Fontaine, G., Green, E. M., Brassard, P., \& Chayer, P. 2008, \aap, 489, 377
\bibitem[Charpinet et al.(2009)]{Charpinet09} Charpinet S., Fontaine, G., \& Brassard, P. 2009, \aap, 493, 595
\bibitem[Cox et al.(1992)]{Cox92} Cox, A. N., Morgan, Siobahn, M., Rogers, F. J., Iglesias, \& C. A. 1992, \apj, 393 272
\bibitem[Dupret et al.(2004)]{Dupret04} Dupret, M.-A., Grigahc\`ene, A., Garrido, R., Gabriel, M., \& Scuflaire, R. 2004, \aap, 414, L17
\bibitem[Dupret et al.(2005)]{Dupret05} Dupret, M.-A., Grigahc\`ene, A., Garrido, R., Gabriel, M., \& Scuflaire, R. 2004, \aap, 435, 927
\bibitem[Dziembowski et al.(1993a)]{Dziembowski93a} Dziembowski, W. A., \& Pamyatnykh, A. A. 1993a, \mnras, 262, 204
\bibitem[Dziembowski et al.(1993b)]{Dziembowski93b} Dziembowski, W. A., Moskalik, P., \& Pamyatnykh, A. A. 1993b, \mnras, 265 588
\bibitem[Gargett \& Ruddick(2003)]{Gargett03} Gargett, A., Ruddick, B., eds. 2003, Double-Diffusion in Oceanography
(Oxford:Pergamon), 381
\bibitem[Grevesse \& Noels(1993)]{Grevesse93} Grevesse, N., \& Noels, A. 1993, in Origin and Evolution of the Elements, Eds Kubono, S. and Kajino, T., 14
\bibitem[Guzik et al.(2000)]{Guzik00} Guzik, J. A., Kaye, A. B., Bradley, P. A., Cox, A. N., \& Neuforge, C. 2000, \apj, 542, L57
\bibitem[Handler et al.(2006)]{Handler06} Handler, G., Jerzykiewicz, M., Rodríguez, E., Uytterhoeven, K., Amado, P. J., Dorokhova, T. N., Dorokhov, N. I., Poretti, E., Sareyan, J.-P., Parrao, L., and 21 coauthors, 2006, \mnras, 365, 327
\bibitem[Hui-Bon-Hoa(2008)]{Hui08} Hui-Bon-Hoa, A. 2008, \apss, 316, 55
\bibitem[Iglesias \& Rogers(1996)]{Iglesias96} Iglesias, C. A., \& Rogers, F. J. 1996, \apj, 464, 943
\bibitem[Jeffery \& Saio(2007)]{Jeffery07} Jeffery, C. S., \& Saio, H. 2007, \mnras, 378, 379
\bibitem[Jerzykiewicz et al.(2005)]{Jerzykiewicz05}Jerzykiewicz, M., Handler, G., Shobbrook, R. R., Pigulski, A., Medupe, R., Mokgwetsi, T., Tlhagwane, P., \& Rodríguez, E. 2005, \mnras, 360, 619
\bibitem[Kippenhahn et al.(1980)]{Kippenhahn80} Kippenhahn, R., Ruschenplatt, G., \& Thomas, H.-C. 1980, \aap, 91, 175
\bibitem[Kiriakidis et al.(1992)]{Kiriakidis92} Kiriakidis, M., El Eid, M. F., \& Glatzel, W.	1992, \mnras, 255, 1
\bibitem[Ko{\l}aczkowski et al.(2006)]{Kolaczkowski06} Ko{\l}aczkowski, Z., Pigulski, A., Soszy{\'n}ski, I., Udalski, A., Kubiak, M., Szyma{\'n}ski, M., {\.Z}ebru{\'n}, K., Pietrzy{\'n}ski, G., Wo{\'z}niak, P. R., Szewczyk, O., \& Wyrzykowski, {\L}. 2006, Memorie della Societa Astronomica Italiana, vol. 77, 336
\bibitem[LeBlanc \& Alecian(2004)]{Leblanc04} LeBlanc, F., \& Alecian, G. 2004, \mnras, 352, 1329
\bibitem[Michaud (1970)]{Michaud70} Michaud, G. 1970, \apj, 160, 641
\bibitem[Miglio et al.(2007a)]{Miglio07a} Miglio, A., Montalb\'an, J., \& Dupret, M.-A. 2007, \mnras, 375, 21
\bibitem[Miglio et al.(2007b)]{Miglio07b} Miglio, A., Bourge, P.-O., Montalb\'an, J., \& Dupret, M.-A. 2007, CoAst, 150, 209
\bibitem[Montalb\'an et al.(2007)]{Montalban07} Montalb\'an, J., Miglio, A., \& Th\'eado, S. 2007, CoAst, 150, 137
\bibitem[Moskalik et al.(1992)]{Moskalik92} Moskalik, P., \& Dziembowski, W. A. 1992, \aap, 256, L5
\bibitem[O'Donoghue et al.(1997)]{Odonoghue97} O'Donoghue, D., Lynas-Gray, A. E., Kilkenny, D., Stobie, R. S., \& Koen, C. 1997, \mnras, 285, 657
\bibitem[Pamyatnykh et al.(2004)]{Pamyatnykh04} Pamyatnykh, A. A., Handler, G., \& Dziembowski, W. A. 2004, \mnras, 350, 1022
\bibitem[Richard et al.(2001)]{Richard01} Richard, O., Michaud, G., \& Richer, J. 2001, \apj, 558, 377
\bibitem[Richard et al.(2004)]{Richard04} Richard, O., Th\'eado, S., \& Vauclair, S. 2004, SoPh, 220, 243
\bibitem[Richer et al.(2000)]{Richer00} Richer, J., Michaud, G., \& Turcotte, S. 2000, \apj, 529, 338 
\bibitem[Rogers \& Iglesias(1992)]{Rogers92} Rogers, F. J., \& Iglesias, C. A. 1992, \apjs, 79, 507
\bibitem[Santos et al.(2001)]{Santos01} Santos, N.C., Israelian, G., \& Mayor, M. 2001, \aap, 373, 1019
\bibitem[Seaton(2005)]{Seaton05} Seaton, M. J. 2005, \mnras, 362, L1
\bibitem[Stancliffe(2009)]{Stancliffe09} Stancliffe, R.J. 2009, \mnras, 394, 1051
\bibitem[Stancliffe et al.(2007)]{Stancliffe07} Stancliffe, R.J., Glebbeek, E., Izzard, R.G., \& Pols, O.R. 2007, \aap, 464, 57
\bibitem[Stancliffe \& Glebbeek(2008)]{Stancliffe08} Stancliffe, R.J., \& Glebbeek, E. 2008, \mnras, 389, 1828
\bibitem[Stankov, A. \& Handler(2005)]{Stankov05} Stankov, A., \& Handler, G. 2005, \apjs, 158, 193
\bibitem[Stothers \& Simon(1969)]{Stothers69} Stothers, R., \& Simon, N.R.  1969, \apj, 157, 673
\bibitem[Th\'eado et al.(2009)]{Theado09} Th\'eado, S, Alecian, G., \& LeBlanc, F. 2009, in prep.
\bibitem[Thompson et al.(2008)]{thompson08} Thompson, I.B., Ivans, I.I, Bisterzo, S., Sneden, C., Gallino, R., Vauclair, S., Burley, G.S., Shectman, S.A., \& Preston, G.W. 2008, \apj, 677, 556
\bibitem[Turcotte et al.(2000)]{Turcotte00} Turcotte, S., Richer, J., Michaud, G., \& Christensen-Dalsgaard, J. 2000, \aap, 360, 603
\bibitem[Ulrich(1972)]{Ulrich72} Ulrich, R. K. 1972, \apj, 172, 165
\bibitem[Vauclair(1975)]{Vauclair75} Vauclair, S. 1975, \aap, 45, 233
\bibitem[Vauclair(2004)]{Vauclair04} Vauclair, S. 2004, \apj, 605, 874


\end{thebibliography}
\end{document}